# The Targets of Opportunity Source Catalog for the EUSO-SPB2 Mission

**Hannah Wistrand,[a,*] Tobias Heibges,[a] Jonatan Posligua,[b] Claire Guépin,[c,d] Mary Hall Reno[b] and Tonia M. Venters[e] for the JEM-EUSO Collaboration**

[a] *Colorado School of Mines, Department of Physics, Golden, CO, USA*

[b] *University of Iowa,Department of Physics and Astronomy, Iowa City, IA, USA*

[c] *University of Chicago, KICP, Chicago, IL, USA*

[d] *Laboratoire Univers et Particules de Montpellier, Montpellier, France*

[e] *Goddard Space Flight Center, Greenbelt, MD, USA*

*E-mail:* hwistrand@mines.edu

The Extreme Universe Space Observatory on a Super Pressure Balloon 2 (EUSO-SPB2) mission was designed to take optical measurements of extensive air showers (EASs) from suborbital space. The EUSO-SPB2 payload includes an optical Cherenkov Telescope (CT), which searches above and below the Earth's limb. Above the limb, the CT measures Cherenkov light from PeV-scale EASs induced by cosmic rays. Below the limb, the CT searches for upwards-going Cherenkov emission from PeV-scale EASs induced by tau neutrinos, to follow up on astrophysical Targets of Opportunity (ToO). Target candidates include gamma-ray bursts, tidal disruption events, and - after the start of the O4 observation run from Ligo-Virgo-Kagra - binary neutron star mergers. Reported here is the selection and prioritization of relevant ToOs from alert networks such as the General Coordinates Network, Transient Name Server, and Astronomer Telegrams, and the translation to a viewing schedule for EUSO-SPB2. EUSO-SPB2 launched on a NASA super pressure balloon in May of 2023 from Wanaka, NZ.

*Speaker

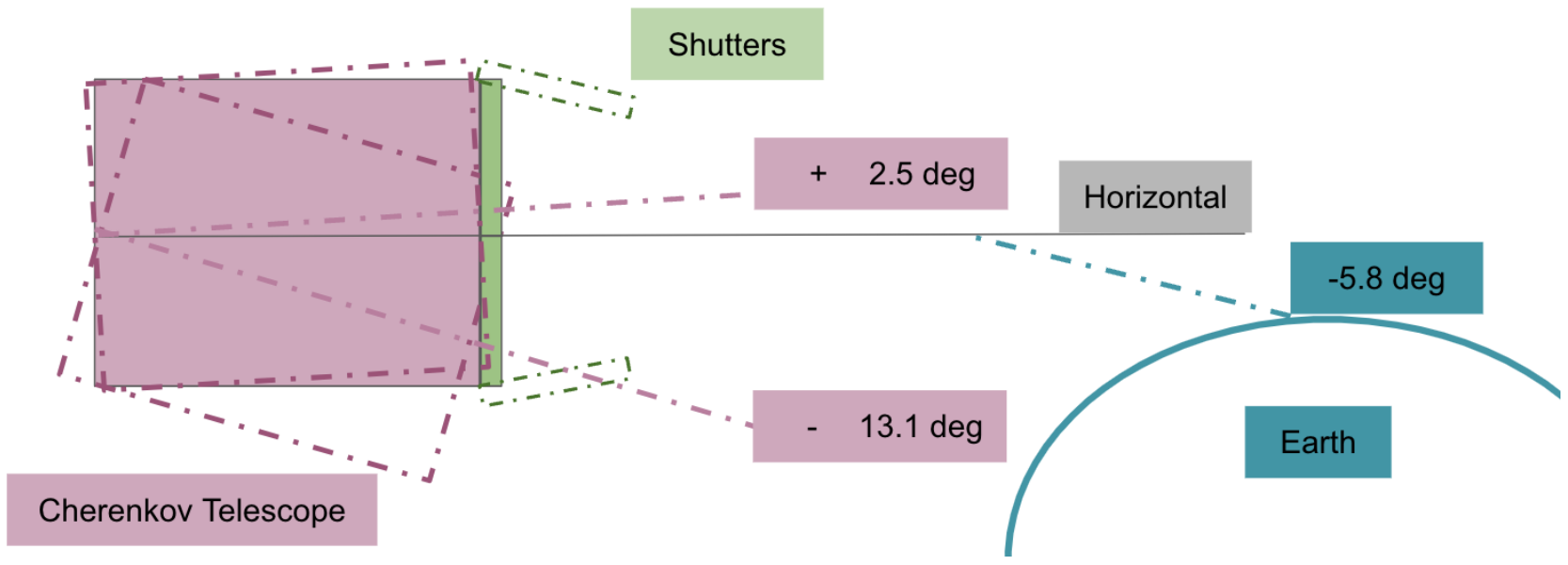

**Figure 1:** Field of view of the EUSO-SPB2 Cherenkov detector. The detector utilizes a tilt and rotation system to achieve a field of view of 6.4° in altitude and 12.8° in azimuth, and a rotational range of 360° in azimuth and a tilting range of 15.6°.

## 1. Introduction

The Extreme Universe Space Observatory on a Super Pressure Balloon 2 (EUSO-SPB2) was developed to observe ultra high energy cosmic rays and very high energy (VHE) neutrinos from suborbital space [1]. Earth-skimming tau-neutrinos can interact with the Earth to produce tau leptons which decay to initiate extensive air showers (EASs). These EASs can then be observed via their optical Cherenkov emissions. IceCube [2] and ANTARES [3] have paved the way for PeV scale searches that rely on large target volumes (through the use of ice or the deep sea) to overcome the challenging detection rate characteristic of neutrinos of this energy scale.

Below PeV energies, ground-based detectors have relied on their sky coverage to increase the likelihood of covering a neutrino source. However, at higher energies, much of the sky becomes inaccessible to ground-based detectors because of Earth attenuation effects. As the interest in these multi-messenger particles increases, the search for alternative detection techniques that are space-based is ongoing. EUSO-SPB2 pioneers a solution to the attenuation problem while overcoming a restricted field of view as a step toward a space-based observatory such as POEMMA [4] with the ability to respond to potential neutrino alerts throughout the sky.

EUSO-SPB2 features two 1m diameter aperture telescopes carried as the payload on a Super Pressure Balloon. The Fluorescence Telescope (FT), which points down, records fluorescence light from cosmic ray EASs with energies above 1 EeV. The Cherenkov Telescope (CT) features a silicon photomultiplier camera system that points near the Earth's limb [5, 6]. Below the limb, the CT follows up on astrophysical transient source alerts and selected steady state sources by searching for upward-going optical Cherenkov emission from PeV-scale EASs induced by tau neutrinos. These transient sources that pass into the CT field of view are called Targets of Opportunity (ToO).

The CT utilizes a tilt and rotation system to follow up on ToOs. The CT achieves a field of view of 6.4° in altitude and 12.8° in azimuth. The entire payload can be rotated 360° in azimuth and the CT can be tilted over a range of 15.6°.

### 1.1 Astrophysical Sources of Interest

The ToO sources of interest for EUSO-SPB2 include galactic and extra-galactic supernovae, binary neutron star mergers, neutron star-black hole mergers, nearby tidal disruption events, flaring blazars, gamma-ray bursts, and other transients (see Fig. 2). It has long been theorized that sources such as these may produce VHE neutrinos, and discoveries such as that of IceCube-170922A in which a neutrino of energy between 0.1-1 PeV was detected in association with a flaring blazar, have further motivated these searches [2].

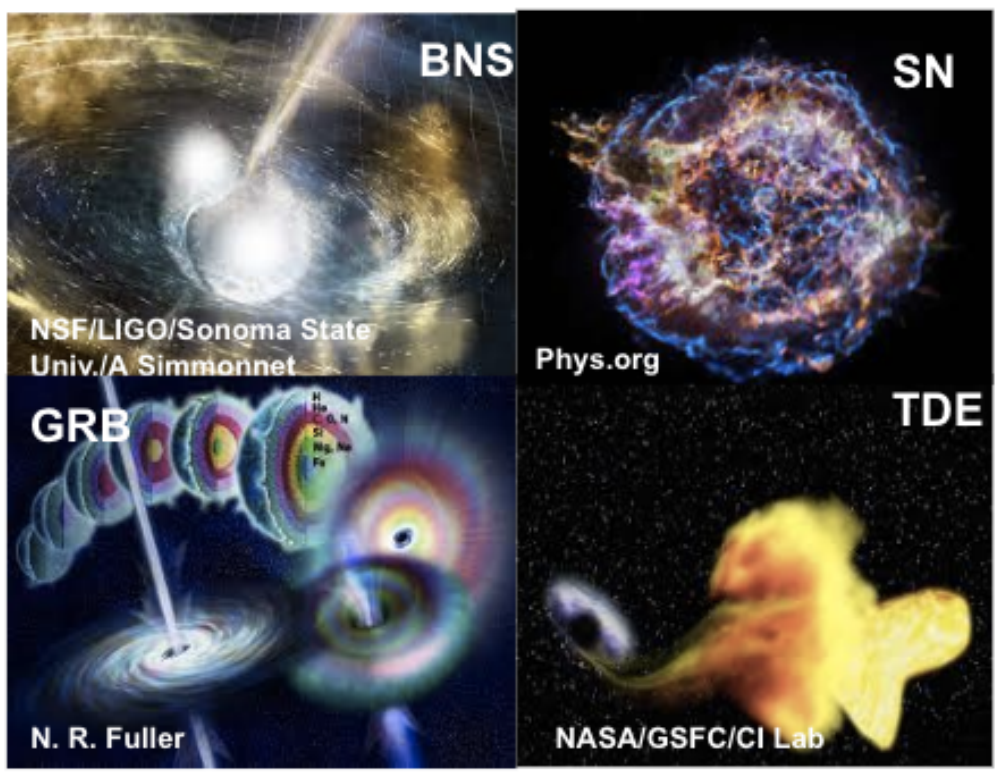

**Figure 2:** Artist representations of some of the types of sources of interest to the EUSO-SPB2 targets of opportunity program.

The transients listed above all offer a chance of neutrino detection to varying degrees. Binary neutron star (BNS) mergers may result in rapidly-spinning magnetars that can accelerate particles with ultra-high energies. These accelerated UHECRs may then produce neutrinos through interactions with the surrounding environment [7, 8]. Tidal disruption events (TDEs) accrete material and produce a flare of radiation that can last from months to years. Some TDEs result in powerful, relativistic jets which may act as proton and nuclei accelerators capable of reaching ultra-high energies and producing VHE and ultra-high energy (UHE) neutrinos [9, 10]. Blazar flares and active galactic nuclei can produce ultra-high energy neutrinos through the interaction of accelerated cosmic rays with photon and baryon backgrounds [11, 12]. Supermassive binary black hole (BBH) mergers may also accelerate still-present material to ultra-high energies which would produce neutrinos through interactions with the surrounding environment, though the potential for these jets is still disputed [7, 13]. Short duration transients, such as gamma-ray bursts (GRBs), may also produce ultra high energy neutrinos [14, 15]. However, due to their short duration and small radiation zone, their strong magnetic fields and radiation backgrounds may limit the maximum energies of transient neutrinos to the EeV range or below [16].

## 2. ToO Catalog for EUSO-SPB2

With the goal of optimizing the scientific reach of the ToO program within the operational parameters of the EUSO-SPB2 mission, a catalog of sources was developed from the three preexisting alert systems broadcast from various observatories and instruments. The ToO catalog is populated through a combination of manual and automated processes that parsed alerts from the Gamma-ray Coordination Network (GCN), the Transient Name Server (TNS), and the Astronomer's Telegram (ATel) are monitored and sorted based on their relation to potential neutrino sources of interest. GCN and TNS alerts are sorted by a python script, and ATel alerts are sorted by hand. The catalog also includes a list of hand-selected steady sources (Fig. 3). For the EUSO-SPB2 observation period on May 14, 2023, the ToO catalog contained 152 sources, including 16 sources from GCN, 12 sources from TNS, 4 sources from ATels, and 120 steady sources.

| | | | |
|---|---|---|---|
| **Steady HBL** | **Steady Hotspot** | **Steady HBL** | **Steady HBL** |
| PKS 0548-322 | Old TA Hotspot | RX J0648 | H 1722+119 |
| (-9.00, 221.58) | (-9.00, 130.12) | (-9.00, 282.20) | (-9.00, 81.59) |
| **Steady HBL** | **Steady Seyfert** | **Steady HBL** | **Steady HBL** |
| 1ES 0347-121 | NGC 253 | HESS J1943-213 | SHBL J001355.9-18540 |
| (-9.00, 249.35) | (-9.00, 130.12) | (-9.00, 70.46) | (-9.00, 120.82) |
| **Steady HBL** | **Steady Blazar** | **Steady HBL** | **Steady HBL** |
| 1ES 1011+496 | Mrk 421 | 1ES 1215+303 | H 2356-309 |
| (-9.00, 324.05) | (-9.00, 310.25 | (-9.00, 301.54) | (-9.00, 135.74 |
| **Steady HBL** | **Steady Blazar** | **Steady HBL** | **Steady HBL** |
| 1ES 0806+524 | Mrk 501 | 1ES 1741+196 | PKS 0301-243 |
| (-9.00, 328.00) | (-9.00, 48.15) | (-9.00, 72.53) | (-9.00, 234.45) |
| **Steady HBL** | **Steady Blazar** | **Steady HBL** | **Steady HBL** |
| RGB J0710+591 | BL Lacertae | 1ES 1727+502 | PKS 0447-439 |
| (-9.00, 335.61) | (-9.00, 44.36) | (-9.00,c34.65) | (-9.00, 199.47) |
| **Steady HBL** | **Steady HBL** | **Steady HBL** | **Steady HBL** |
| 1ES 0414+009 | PKS 2155-304 | KUV 00311-1938 | 1ES 1218+304 |
| (-9.00, 265.72) | (-9.00, 135.28) | (-9.00, 123.38) | (-9.00, 302.11) |

**Table 1:** List of several steady sources included in the ToO source catalog which were visible for the EUSO-SPB2 observation period on May 14, 2023. These sources are never removed from the catalog. Each catalog entry includes the source name, type (if known), the original observation time, and the original observation coordinates.

Each catalog (ToO) entry includes the reported observation time and observation coordinates of the alert. If the information is available via the respective alert system, the redshift and other characteristics are calculated and included in the catalog entry. Next, each source is assigned a priority value. These priority values are assigned based on the scheme presented in Section 3. From this information, the time each ToO enters the detector's field of view on a given night can then be calculated. ToO observations are scheduled for each nightly period of possible CT operation based on the values included in their catalog entries. More information on this scheduling process can be found in [17, 18].

A user interface that was developed for the ToO mission, so that shifters could manually enter new source alert information at the time of observation if an alert of high priority was issued during the current observation run. This manual process calculates the time and CT pointing direction needed to catch the source and adds the information to the ToO observing schedule for the current observation run.

## 3. ToO Prioritization for EUSO-SPB2

The mission source catalog is constantly updated to add selected new alerts and to remove outdated alerts. It typically contains more than 150 sources. On a given night, it is possible for more than 10 of these sources to fulfill observability criteria. Because observation time is limited each night, prioritization of these ToOs is required to ensure the most promising sources are viewed by the CT during the observation run. The process to develop the EUSO-SPB2 prioritization scheme considered existing HE-VHE neutrino source models, coincidence studies, the rate of occurrence for each ToO source type, and the age of the source. The final prioritization scheme is shown in Table 2 and summarized below.

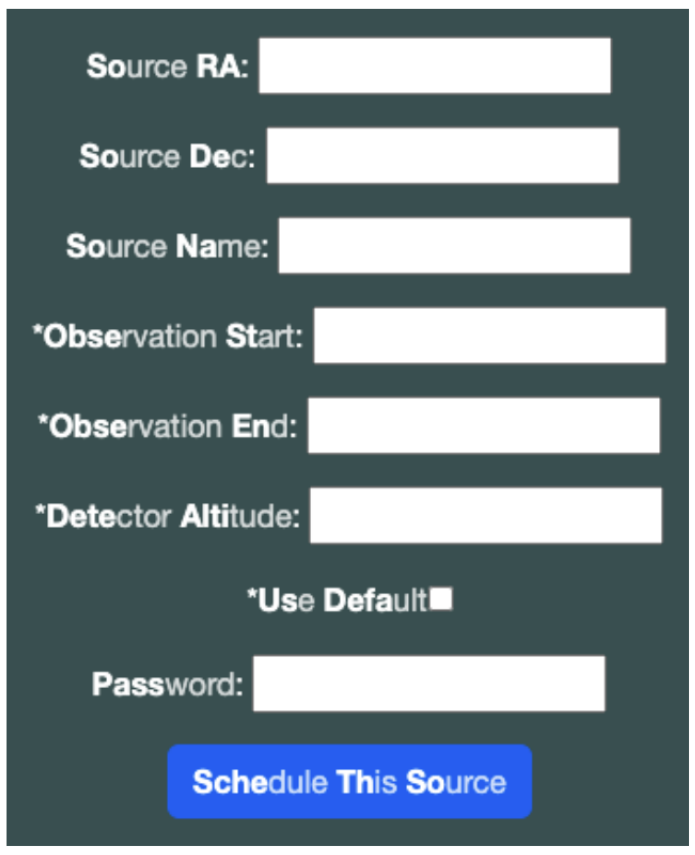

**Figure 3:** The form on the user interface to add a new source to the ToO source catalog, then to rerun the scheduler for the current observation period with the updated catalog.

| Source Type | EUSO-SPB2 Priority Tier |
|---|---|
| Galactic supernovae | 1 |
| Binary neutron star and black hole-neutron star mergers | 2 |
| Nearby tidal disruption events | 3 |
| Flaring blazar or active galactic nuclei | 4 |
| Gamma-ray bursts | 5 |
| Supernovae outside of the galaxy | 6 |
| Other transients | 7 |
| Steady sources | 8 |

**Table 2:** The list of ToO sources of interest and their respective priority values as assigned for the EUSO-SPB2 mission.

**Galactic Supernovae:** Supernovae are widely considered to be the strongest and most frequent source of cosmic neutrinos. While extremely rare, galactic supernovae are expected to produce the highest rate of detectable neutrinos. These characteristics earn the top priority spot in the ToO prioritization scheme [7, 19].

**Binary neutron star and black hole-neutron star mergers:** The expected neutrino production rate of these events is contested, though it is expected that not every event would result in a stable enough environment to produce neutrinos. Because of the theoretical nature of these sources' relationship to neutrino production, binary neutron star and black hole-neutron star mergers have been placed at a lower priority. However, because of the rare nature of these events as well as the promising LIGO-Virgo reports of candidate electromagnetic counterparts, these events are extremely fascinating and have thus been awarded the number two priority spot [7, 8].

**Nearby tidal disruption events:** As explained in Section 1.1, TDEs may result in powerful,

relativistic jets which may produce very-high and ultra-high energy neutrinos [9]. Long duration TDEs are especially promising for UHE neutrino production, but may be subject to energy losses. However, very few nearby TDEs have produced such jets. This conditional promise of UHE neutrinos in addition to the rare nature of nearby jetting TDEs earns the third priority spot [16].

**Flaring blazar or active galactic nuclei:** Neutrino detection has already coincided with blazar flare events, as evidenced by IceCube-170922A in association with a flaring blazar, TXS 0506+056. Similar success is expected with other types of active galactic nuclei. However, these events are likely to produce neutrinos at sub-PeV energies and are therefore given a lower priority [2, 11, 12].

**Gamma-ray bursts:** Gamma-ray bursts are a type of short-duration transient. With a similar promise of high energy neutrinos as the TDEs, this event type is placed at a much lower priority due to its short duration and small radiation zone which may limit the maximum energy of neutrinos to the EeV range or below - much higher than the PeV range of the CT neutrino energy threshold [16].

**Supernovae outside of the galaxy:** Like galactic supernovae, supernovae outside the galaxy are widely considered to be the strongest and most frequent source of cosmic neutrinos. Unlike galactic supernovae, these events are only expected to produce tens of detectable neutrinos. This, despite the higher rate of occurrence, places the extragalactic supernova low on the priority list [7, 19].

**Other transients:** Due to their high power, transient sources are expected to supply a significant number of detectable neutrinos. Because of the specific expectations of the sources listed above, they have been included in the ToO prioritiztion scheme in a specific order. While all other transients are expected to provide a promising opportunity to detect neutrinos, the lack of knowledge and specificity of this category places them low on the priority list [7, 16].

**Steady sources:** Steady sources are much less time-dependent in their observation opportunities. While, for example, short-duration transients may only yield promising results for less than 5 days. It is this reason that the list of steady sources included in the ToO catalog is prioritized last. Because of this, all other sources will be scheduled before the steady sources, with the steady sources only being scheduled if there is remaining observation time.

## 4. ToO Scheduling

A prioritization scheme is critical for the ToO scheduling process. The scheduling process, which is outlined in [7], takes entries from the ToO catalog, compares their daily trajectories with the CT trajectory, and calculates the time and location of when the ToO will cross the detector's field of view. After this is finished, there can be tens of sources that cross into the detector's field of view during a nightly observation run. Because each of these runs may only be 4-5 hours at the most, with each ToO being visible anywhere from 20 minutes to 1 hour and 20 minutes, 4 or 5 sources at most may be scheduled (see Fig. 4). The prioritization scheme is used to trim down the observable ToOs to fit the observation run schedule.

For a given observation period, if the number of observable sources is more than 5, the ToO software parses through the observable sources and chooses 4-5 based on priority values. This process involves 7 rounds of scheduling. Each round the software searches for a new priority value. If a transient source (the first 7 priority ratings) is observable with that priority value, it gets scheduled. If not, the program moves onto the next priority value. This continues until 4-5 sources are scheduled. If after 7 rounds there remains time in the observation period, the process starts again. The software will continue to fill the available time until each observable source has been checked. If there are not enough observable transient sources to fill the schedule, the program fills the schedule with observable steady sources (priority 8 sources).

| Source type, name | Pointing (az) | Move time | Start time | End time |
|---|---|---|---|---|
| AGN, SDSS J102906.69+555625.2 | 47.19° | 05:00 | 05:10 | 05:30 |
| Steady FRB, FRB 20181119A | 6.23° | 05:30 | 05:40 | 06:10 |
| SN II, SN 2023ftg | 81.76° | 06:10 | 06:20 | 06:50 |
| AGN, 4FGL J0910.0+4257 | 315.96° | 07:50 | 08:00 | 08:40 |
| GRB, GRB230503A | 187.66° | 09:10 | 09:20 | 10:40 |

**Table 3:** Schedule produced by the software for May 14, 2023. For each source, we show the CT pointing, the move, start and end times in UTC. For the pointing we only indicate the azimuth (in degrees), as in the VHE neutrino observation mode the altitude is fixed to $-9°$.

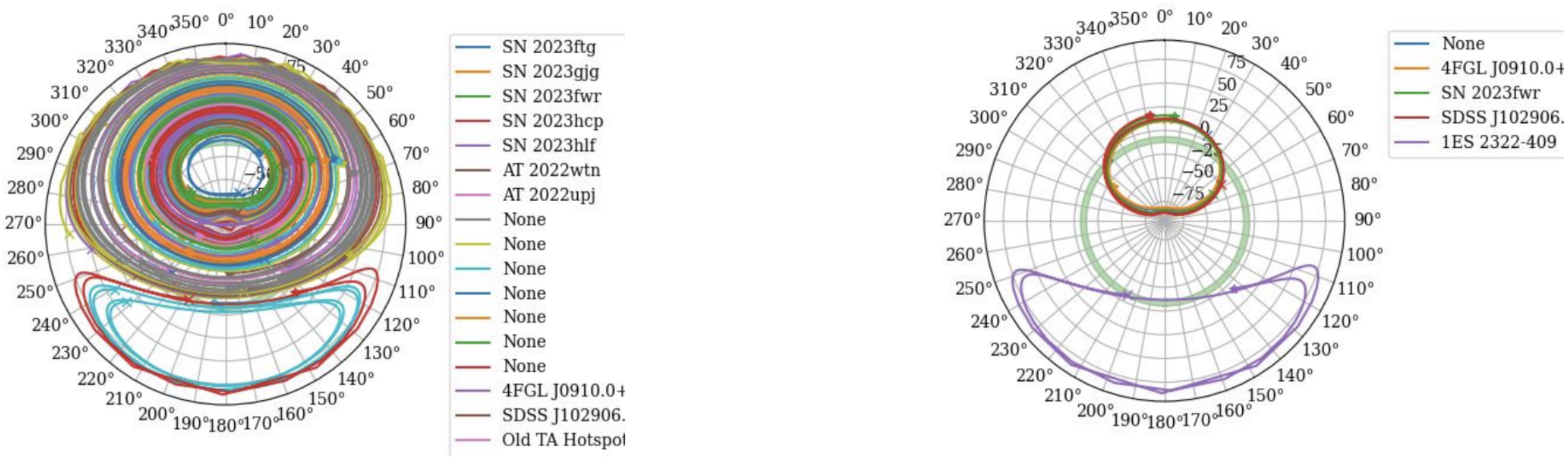

**Figure 4:** On the left: An example of the trajectories of ToOs in the EUSO-SPB2 Cherenkov detector field of view before prioritization on May 14, 2023. There are approximately 70 ToOs that come into the field of view during this time - many more than can reasonably be observed during a single observation run. On the right: the trajectories of ToOs in the EUSO-SPB2 Cherenkov telescope field of view after prioritization on May 14, 2023. The number of sources has been reduced to 5 for the average 5 hour observation window. The green circle on each of these plots represents the field of view, and the radius represents the altitude.

If there are multiple observable transient sources of a given priority value, the software checks a database of how much time the sources have already been observed. This database is updated after every observation run and records the observed sources with the amount of time (in minutes) of the ToO observation. Sources that have less recorded previous observation time in the database are prioritized over sources that have already been observed. An example schedule from May 14, 2024 is included in Table 3.

## 5. Conclusion

The EUSO-SPB2 flew with a live prioritized catalog of potential neutrino sources to inform CT pointing. The ToO source catalog with its prioritization scheme was optimized to reduce the guesswork from the observation schedule of the EUSO-SPB2 CT and to optimize the neutrino-mode observation time. This process offers a useful application of the many alert systems from successful detectors, and it can be used for various missions aimed at observing VHE neutrinos - including a follow-up balloon mission to EUSO-SPB2. Applications include identifying potential observing targets for Mini-EUSO [20], a JEM-EUSO instrument equipped with a fluorescence telescope installed on the International Space Station. Additional applications include other balloon missions such as PUEO [21] and satellite missions such as Terzina [22] and POEMMA [4]. After the software is finalized and documented, it will be open source and available via GitHub.

**Acknowledgements** – The authors acknowledge the support by NASA awards 11-APRA-0058, 16-APROBES16-0023, 17-APRA17-0066, NNX17AJ82G, NNX13AH54G, 80NSSC18K0246, 80NSSC18K0473,80NSSC19K0626, 80NSSC18K0464, 80NSSC22K1488, 80NSSC19K0627 and 80NSSC22K0426, the French space agency CNES, National Science Centre in Poland grant n. 2017/27/B/ST9/02162, and by ASI-INFN agreement n. 2021-8-HH.0 and its amendments. This research used resources of the US National Energy Research Scientific Computing Center (NERSC), the DOE Science User Facility operated under Contract No. DE-AC02-05CH11231. We acknowledge the NASA BPO and CSBF staffs for their extensive support. We also acknowledge the invaluable contributions of the administrative and technical staffs at our home institutions.

# Full Authors list: The JEM-EUSO Collaboration

S. Abe[ff], J.H. Adams Jr.[ld], D. Allard[cb], P. Alldredge[ld], R. Aloisio[ep], L. Anchordoqui[le], A. Anzalone[ed,eh], E. Arnone[ek,el], M. Bagheri[lh], B. Baret[cb], D. Barghini[ek,el,em], M. Battisti[cb,ek,el], R. Bellotti[ea,eb], A.A. Belov[ib], M. Bertaina[ek,el], P.F. Bertone[lf], M. Bianciotto[ek,el], F. Bisconti[ei], C. Blaksley[fg], S. Blin-Bondil[cb], K. Bolmgren[ja], S. Briz[lb], J. Burton[ld], F. Cafagna[ea.eb], G. Cambié[ei,ej], D. Campana[ef], F. Capel[db], R. Caruso[ec,ed], M. Casolino[ei,ej,fg], C. Cassardo[ek,el], A. Castellina[ek,em], K. Černý[ba], M.J. Christl[lf], R. Colalillo[ef,eg], L. Conti[ei,en], G. Cotto[ek,el], H.J. Crawford[la], R. Cremonini[el], A. Creusot[cb], A. Cummings[lm], A. de Castro Gónzalez[lb], C. de la Taille[ca], R. Diesing[lb], P. Dinaucourt[ca], A. Di Nola[eg], T. Ebisuzaki[fg], J. Eser[lb], F. Fenu[eo], S. Ferrarese[ek,el], G. Filippatos[lc], W.W. Finch[lc], F. Flaminio[eg], C. Fornaro[ei,en], D. Fuehne[lc], C. Fuglesang[ja], M. Fukushima[fa], S. Gadamsetty[lh], D. Gardiol[ek,em], G.K. Garipov[ib], E. Gazda[lh], A. Golzio[el], F. Guarino[ef,eg], C. Guépin[lb], A. Haungs[da], T. Heibges[lc], F. Isgrò[ef,eg], E.G. Judd[la], F. Kajino[fb], I. Kaneko[fg], S.-W. Kim[ga], P.A. Klimov[ib], J.F. Krizmanic[lj], V. Kungel[lc], E. Kuznetsov[ld], F. López Martínez[lb], D. Mandát[bb], M. Manfrin[ek,el], A. Marcelli[ej], L. Marcelli[ei], W. Marszał[ha], J.N. Matthews[lg], M. Mese[ef,eg], S.S. Meyer[lb], J. Mimouni[ab], H. Miyamoto[ek,el,ep], Y. Mizumoto[fd], A. Monaco[ea,eb], S. Nagataki[fg], J.M. Nachtman[li], D. Naumov[ia], A. Neronov[cb], T. Nonaka[fa], T. Ogawa[fg], S. Ogio[fa], H. Ohmori[fg], A.V. Olinto[lb], Y. Onel[li], G. Osteria[ef], A.N. Otte[lh], A. Pagliaro[ed,eh], B. Panico[ef,eg], E. Parizot[cb,cc], I.H. Park[gb], T. Paul[le], M. Pech[bb], F. Perfetto[ef], P. Picozza[ei,ej], L.W. Piotrowski[hb], Z. Plebaniak[ei,ej], J. Posligua[li], M. Potts[lh], R. Prevete[ef,eg], G. Prévôt[cb], M. Przybylak[ha], E. Reali[ei,ej], P. Reardon[ld], M.H. Reno[li], M. Ricci[ee], O.F. Romero Matamala[lh], G. Romoli[ei,ej], H. Sagawa[fa], N. Sakaki[fg], O.A. Saprykin[ic], F. Sarazin[lc], M. Sato[fe], P. Schovánek[bb], V. Scotti[ef,eg], S. Selmane[cb], S.A. Sharakin[ib], K. Shinozaki[ha], S. Stepanoff[lh], J.F. Soriano[le], J. Szabelski[ha], N. Tajima[fg], T. Tajima[fg], Y. Takahashi[fe], M. Takeda[fa], Y. Takizawa[fg], S.B. Thomas[lg], L.G. Tkachev[ia], T. Tomida[fc], S. Toscano[ka], M. Traïche[aa], D. Trofimov[cb,ib], K. Tsuno[fg], P. Vallania[ek,em], L. Valore[ef,eg], T.M. Venters[lj], C. Vigorito[ek,el], M. Vrabel[ha], S. Wada[fg], J. Watts Jr.[ld], L. Wiencke[lc], D. Winn[lk], H. Wistrand[lc], I.V. Yashin[ib], R. Young[lf], M.Yu. Zotov[ib].

[aa] Centre for Development of Advanced Technologies (CDTA), Algiers, Algeria

[ab] Lab. of Math. and Sub-Atomic Phys. (LPMPS), Univ. Constantine I, Constantine, Algeria

[ba] Joint Laboratory of Optics, Faculty of Science, Palacký University, Olomouc, Czech Republic

[bb] Institute of Physics of the Czech Academy of Sciences, Prague, Czech Republic

[ca] Omega, Ecole Polytechnique, CNRS/IN2P3, Palaiseau, France

[cb] Université de Paris, CNRS, AstroParticule et Cosmologie, F-75013 Paris, France

[cc] Institut Universitaire de France (IUF), France

[da] Karlsruhe Institute of Technology (KIT), Germany

[db] Max Planck Institute for Physics, Munich, Germany

[ea] Istituto Nazionale di Fisica Nucleare - Sezione di Bari, Italy

[eb] Università degli Studi di Bari Aldo Moro, Italy

[ec] Dipartimento di Fisica e Astronomia "Ettore Majorana", Università di Catania, Italy

[ed] Istituto Nazionale di Fisica Nucleare - Sezione di Catania, Italy

[ee] Istituto Nazionale di Fisica Nucleare - Laboratori Nazionali di Frascati, Italy

[ef] Istituto Nazionale di Fisica Nucleare - Sezione di Napoli, Italy

[eg] Università di Napoli Federico II - Dipartimento di Fisica "Ettore Pancini", Italy

[eh] INAF - Istituto di Astrofisica Spaziale e Fisica Cosmica di Palermo, Italy
[ei] Istituto Nazionale di Fisica Nucleare - Sezione di Roma Tor Vergata, Italy
[ej] Università di Roma Tor Vergata - Dipartimento di Fisica, Roma, Italy
[ek] Istituto Nazionale di Fisica Nucleare - Sezione di Torino, Italy
[el] Dipartimento di Fisica, Università di Torino, Italy
[em] Osservatorio Astrofisico di Torino, Istituto Nazionale di Astrofisica, Italy
[en] Uninettuno University, Rome, Italy
[eo] Agenzia Spaziale Italiana, Via del Politecnico, 00133, Roma, Italy
[ep] Gran Sasso Science Institute, L'Aquila, Italy
[fa] Institute for Cosmic Ray Research, University of Tokyo, Kashiwa, Japan
[fb] Konan University, Kobe, Japan
[fc] Shinshu University, Nagano, Japan
[fd] National Astronomical Observatory, Mitaka, Japan
[fe] Hokkaido University, Sapporo, Japan
[ff] Nihon University Chiyoda, Tokyo, Japan
[fg] RIKEN, Wako, Japan
[ga] Korea Astronomy and Space Science Institute
[gb] Sungkyunkwan University, Seoul, Republic of Korea
[ha] National Centre for Nuclear Research, Otwock, Poland
[hb] Faculty of Physics, University of Warsaw, Poland
[ia] Joint Institute for Nuclear Research, Dubna, Russia
[ib] Skobeltsyn Institute of Nuclear Physics, Lomonosov Moscow State University, Russia
[ic] Space Regatta Consortium, Korolev, Russia
[ja] KTH Royal Institute of Technology, Stockholm, Sweden
[ka] ISDC Data Centre for Astrophysics, Versoix, Switzerland
[la] Space Science Laboratory, University of California, Berkeley, CA, USA
[lb] University of Chicago, IL, USA
[lc] Colorado School of Mines, Golden, CO, USA
[ld] University of Alabama in Huntsville, Huntsville, AL, USA
[le] Lehman College, City University of New York (CUNY), NY, USA
[lf] NASA Marshall Space Flight Center, Huntsville, AL, USA
[lg] University of Utah, Salt Lake City, UT, USA
[lh] Georgia Institute of Technology, USA
[li] University of Iowa, Iowa City, IA, USA
[lj] NASA Goddard Space Flight Center, Greenbelt, MD, USA
[lk] Fairfield University, Fairfield, CT, USA
[ll] Department of Physics and Astronomy, University of California, Irvine, USA
[lm] Pennsylvania State University, PA, USA